\documentclass[aps,prb,twocolumn,showpacs,preprintnumbers]{revtex4}

\usepackage{graphicx}
\usepackage{amsmath}

\providecommand\epl{Europhys.\ Lett.\ }
\providecommand\jap{J.\ Appl.\ Phys.\ }
\providecommand\jltp{J.\ Low Temp.\ Phys.\ }
\providecommand\phb{Physica B }
\providecommand\prx{Phys.\ Rev.\ }
\providecommand\rsi{Rev.\ Sci.\ Instrum.\ }

\begin{document}
    

\title{Electrometry on charge traps with a single--electron transistor}

\author{Miha Furlan}
\altaffiliation[Present address: ]{Laboratory for Astrophysics,
        Paul Scherrer Institute, CH-5232 Villigen PSI, Switzerland.}
\affiliation{Solid State Physics Laboratory, ETH Z\"{u}rich,
        CH-8093 Z\"{u}rich, Switzerland\\
        Swiss Federal Office of Metrology and Accreditation METAS,
        CH-3003 Bern-Wabern, Switzerland}
\author{Sergey V.\ Lotkhov}
\affiliation{Physikalisch-Technische Bundesanstalt PTB,
        D-38116 Braunschweig, Germany}

\date{\today}

\begin{abstract}
Background charge fluctuators are studied individually by means of a
modified single--electron pump. Operation of the device in a feedback mode
allows electrometric sensing of the charged background and its behavior
upon electric potential variations due to geometrically different gates.
Pulse height spectra and hysteresis of charge trapping transitions are
discussed as a specific signature of distinct fluctuators. The location of
individual traps is determined from the experimental data and based on
electrostatic calculations.
\end{abstract}

\pacs{73.23.Hk, 71.55.-i, 72.20.Jv}

\maketitle

\section{\label{sec:intro}Introduction}
Single--electron charging effects are among the most celebrated
phenomena of mesoscopic physics. The technological trend in continuous
down-scaling of feature size in solid state electronics advances into a field
where mesoscopic and quantum phenomena start to play a dominant role. The
potential advantages of nanostructured electronics are e.g.\ increased
speed, reduced power dissipation or the implementation of quantum logic
operations. However, new attractive phenomena are often accompanied by less
desirable effects. One of the most serious problems of mesoscopic electronic
devices (and an extension of their high sensitivity)
is fluctuating background charge. In spite of increased efforts in
fabrication technology to produce high purity components, imperfections in solid
state devices leading to interface or bulk trap states are responsible for
inevitable background charge fluctuations. The significance of this problem is
well known e.g.\ by the MOSFET community,~\cite{flicker.r} but it is at least as
crucial for potentially useful nanodevice applications.

The metallic single--electron transistor (SET)~\cite{setrev.r,first.r} is one of
the simplest and most extensively studied architectures manifesting Coulomb
blockade. Due to its extreme charge sensitivity and the ability to transfer
individual electrons, SETs have been proposed for a variety of applications or
as building blocks of future electronics. It is known, however, that
device sensitivity at dc operation is limited by low frequency input noise due
to dynamic charge fluctuations in the vicinity of the SET, in
metallic~\cite{setnoise.r,tunneltrap.r} as well as in
semiconductor~\cite{dotnoise.r,trapimaging.r} devices.
A possible solution
to this problem is provided e.g.\ by the radio frequency
SET~\cite{rfset.r} operating well above the $1/f$
noise spectrum, or by appropriate device fabrication aiming for optimal shielding
of the SET island against external charge perturbations.~\cite{stacked.r}

At least as serious as the dynamic noise is the static offset charge problem,
which sets the device in an initial random state due to the random
configuration of the quasi-static charge background. Fluctuators with large
time constants but strong electrostatic
influence on the device characteristics may
shift the operation point such that repeated tuning of the biasing condition is
required. This unpleasant though not unusual scenario severely reduces the
reliability of SETs for broader applications. Furthermore, it has also been
suggested that photon assisted tunneling (PAT) due to background charge
fluctuations as extrapolated from $1/f$ noise spectra may be the
accuracy limiting process in high precision experiments.~\cite{patnoise.r}

From a different perspective, charging effects in ultra small electronic devices
may serve as the base mechanism of memory devices.~\cite{yano.r} Due to
technological limitations, operation of lithographically prepared
single--electron devices is restricted to very low temperatures. However, the
small size of charge traps (on the order of 1~nm) due to intrinsic impurities
potentially allows manipulation of single charges at room temperature. If the
random nature of impurities can be controlled or at least their
behavior selectively discriminated, the discrete charging of traps can provide
the essential process in high-density nonvolatile memory devices.

Therefore, due to the crucial role of background charge fluctuations for
mesoscopic devices (as well as for microelectronics in general), a profound
understanding of charge trapping processes is most desirable. We have performed
extensive experimental studies on charge fluctuations in the vicinity of a
modified multiple gate SET (originally an electron pump, see
section~\ref{sec:samples}). We are able to identify the location and electronic
configuration of individual metastable trap states. Our method proves the
feasibility of and provides a tool for characterizing the charge background and
for optimizing the operation point with a working device with respect to noise
performance, i.e.\ the mesoscopic sample under investigation and the detector
for charge fluctuations are the same.

\section{\label{sec:experiment}Experiment}
Measurements on experimental sensing of environmental charges are the main
focus of this paper. Because the results certainly  depend on the
device topology and the substrate and lithographic quality, a detailed
description of the device fabrication as given in the following part is believed
to be essential. The experimental setup and the measurement procedure, which
allow a virtually continuous resolution of charge fluctuations as a function of
gate potentials, follow in the second part of this section. The third part
presents the results.

\subsection{\label{sec:samples}Device fabrication}
\begin{figure}[h]
\includegraphics[width=8cm]{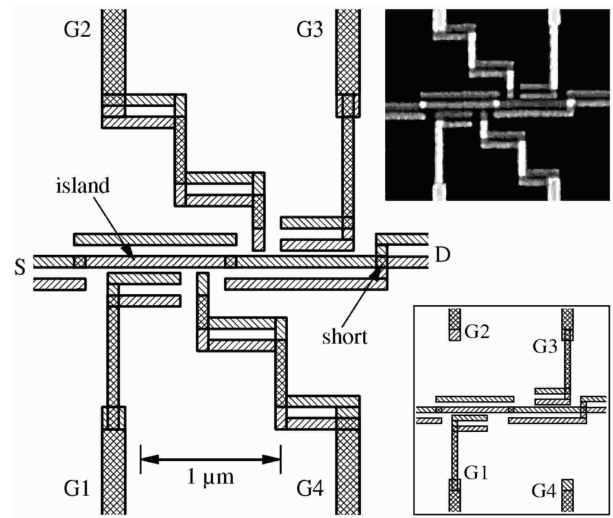}
\caption{\label{layout} Layout of the multigate SET (on the left) showing the
actual metal depositions from the two angle evaporation (note the different
hatching). The island closer to the source (S) couples to the electrodes via
small high resistance tunnel junctions, whereas the other island of the original
e-pump design has a short to drain (D). The AFM picture (inset top right) nicely
shows the overlap regions (brighter is higher) of the same device (type A). An
alternative device used (type B) differs only by more distant gates G2 and G4
(inset bottom right). The narrow strip lines have a width of 80~nm.}
\end{figure}

The investigation of a multigate SET transistor was intended to be a preparation
step towards the study of an electron pump: a device for pumping of single
electrons in a controllable way, which in the simplest version consists of three
ultra small tunnel junctions with two gates.~\cite{pump.r} The correlated
transfer of electrons is realized in the pump with the help of an external
rf-drive applied to the gates with  appropriate relative phase shifts. Since the
electrostatic interactions are generally long-range, it is important for proper
pumping to be able to minimize the inevitable cross-influence of the gates. In
order to experimentally study different gate configurations we used the
multigate SET  sharing the same layout with the pump, with the
exception that one side tunnel junction in the pump was intentionally shorted,
see Fig.~\ref{layout}.

The structure of type Al/AlO$_{\text{x}}$/Al was fabricated by the two angle
evaporation technique~\cite{twoangle.r} on a thermally oxidized (300~nm deep)
wafer of monocrystal Si. The trilayer mask PMMA/Ge/PMMA-MAA (Copolymer), 100~nm,
30~nm and 300~nm thick respectively (top to bottom), was exposed in the e-beam
pattern generator Philips EBPG 4, both in the part of the fine structure and in
the part of the contact leads and pads. After development of the PMMA in the
mixture of MIBK/IPA, 1:2, for 1~min, the pattern was transferred to the layer of
Ge by means of reactive-sputter etching (CF$_{\text{4}}$, 1~Pa, rf-power density
0.2~W/cm$^{2}$, 75~s), and then through the layer of the Copolymer
(O$_{\text{2}}$, 0.8~Pa, 0.1~W/cm$^{2}$, 3~min) down to the substrate oxide with
sufficient over-etch time, which was necessary to compensate for the process
non-uniformities. The undercut space necessary for oblique deposition was formed
by isotropic etching of Copolymer at high oxygen pressure (30~Pa,
0.1~W/cm$^{2}$, 20~min). At this stage the surface oxide of the substrate was
exposed to the oxygen plasma, which we believe might have helped in restoring
the oxide quality after the previous sputter-etch step. The metal structure was
deposited by the e-beam evaporator in one vacuum run and was formed by two
conform layers of Al (25~nm and 35~nm thick) evaporated at 0.3~nm/s at the
angles of $+14^{\circ}$ and $-14^{\circ}$, respectively. The base pressure in
the evaporation chamber was below $10^{-5}$~Pa. The junctions
(80~nm~$\times$~80~nm in size) were formed as the overlapped tips of the Al
microstrips belonging to different layers. The dielectric barrier was formed by
room temperature oxidation of the first Al layer (10~min at pure oxygen of
1~mbar) before the deposition of the second layer. After evaporation the sample
was immersed in acetone for mask lift-off.

The only unknown or uncontrollable process influencing the physical properties
of the samples is the long-term oxidation when stocking them. In spite of
the high-quality fabrication and proven long-term stability of the devices, very
slow oxide formation at the interfaces inevitably continues, which may,
eventually together with ion diffusion, subsequently introduce shallow defects.

\subsection{\label{sec:measure}Experimental setup and measurement procedure}
Measurements were performed in a dilution cryostat with a base temperature of
5~mK, whereas the effective electron temperature of our system was
found~\cite{teff.r} to be 45~mK. The lines running from the room temperature
electronics to the cold SET devices were filtered against microwave propagation
with 2~m Thermocoax cables~\cite{thermocoax.r} each, and the sample holder was
thoroughly enclosed to be electromagnetically tight. A magnetic field of
typically 1~T was applied to suppress superconductivity of the Aluminum devices.

\begin{figure}[h]
\includegraphics[width=7.5cm]{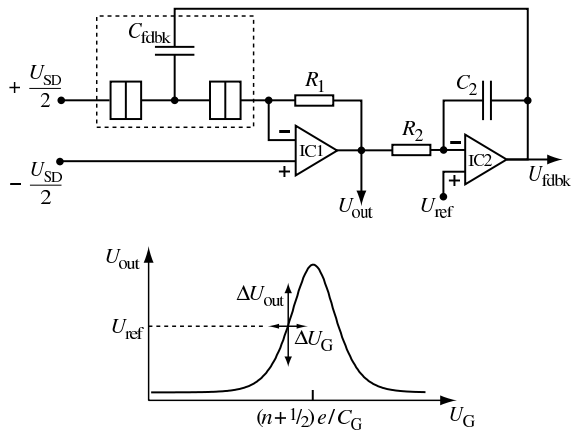}
\caption{\label{readout} Schematic readout circuit with voltage biased SET
(dashed rectangle at low temperature), transimpedance amplifier (IC1) and
feedback circuitry (IC2). Signal attenuators and filters used are omitted here
for clarity. A measured Coulomb blockade peak (non-zero
conductance around $(n + 1/2)e$ island charge) is shown in the lower part,
illustrating the effect of a potential variation $\Delta U_\text{G}$.}
\end{figure}

The bias and readout electronics is schematically drawn in
Fig.~\ref{readout}. It consists of a symmetrically voltage driven SET,
where the current $I_\text{SET}$ through the transistor is sensed by
the transimpedance amplifier IC1 (with feedback resistor $R_1 \gg
R_\text{SET}$, where $R_\text{SET}$ is the asymptotic SET device
series resistance). The input impedance of IC1 is required to be very high, then
the current through the SET is given by
\begin{equation}
I_\text{SET} = -\ \frac{2 U_\text{out} + U_\text{SD}}{2 R_1},
\end{equation}
where $U_\text{SD}$ is the source-drain voltage.

The measurements discussed here were performed with a feedback circuitry holding
the SET island at constant potential (or charge) irrespective of external
potential variations. This was achieved by biasing the device at a Coulomb
blockade peak at high gain $\eta = \text{d} U_\text{out} / \text{d} U_\text{G}$
(see lower part in Fig.~\ref{readout}), integrating $U_\text{out}$ (right hand
 part of circuit with IC2 in Fig.~\ref{readout}) and feeding the signal back to
G$_\text{fdbk}$, the feedback gate. When $U_\text{out}$ changes by $\Delta
U_\text{out}$ for whatever reason, the feedback circuit reacts with a
voltage $\Delta U_\text{fdbk} = - \Delta U_\text{out} / \eta$ in order to adjust
$U_\text{out}$ to $U_\text{ref}$ (the minus sign stems from the inverting
integrator circuit). In general, the requirement for the island to stay on
constant potential is
\begin{equation}
\sum q_{i} = \sum U_{i} C_{i} = constant,
\end{equation}
where $q_{i}$ is the charge influenced on the island by the \mbox{$i$-th} gate with
capacitance $C_{i}$ being at the potential $U_{i}$. Therefore, upon a variation
$\Delta U_{i}$ the island charge is modulated (i.e.\ the gate modulation curve
is shifted by $- \Delta q_{i}/C_{i}$) and the feedback circuit response,
still without background charges, is a voltage

\begin{equation}
\Delta U_\text{fdbk} = - \Delta U_{i} \frac{C_{i}}{C_\text{fdbk}}
\label{eq:fdbk}
\end{equation}
to the feedback gate with  capacitance $C_\text{fdbk}$.
If the device is an isolated system where the only non-zero potentials are
supplied by the  gates, the feedback voltage is always linear to the gate
voltages, see  Eq.~(\ref{eq:fdbk}), and can be subtracted in the data analysis
of gate sweep measurements. However, if extra background charges in the vicinity
of the SET are taken into account, they modify the island potential such that

\begin{equation}
\delta Q_\text{island} = - \Delta U_\text{fdbk} C_\text{fdbk}
\end{equation}
is the corresponding charge shift which is compensated by G$_\text{fdbk}$.

The dynamics of our electronics are limited to low frequencies. The main
constraint on the response is given by the $RC$ time of the SET resistance
($R_\text{SET} = 10^{5} \dots 10^{6} \Omega$) and the cable capacitance
($\approx 500$~pF), constituting roughly a 10~kHz low-pass. The integration time
$R_2 C_2$ was typically set much slower, i.e.\ on the order of 100~ms, to
prevent the system from oscillations. The quasi dc measurements were performed
by scanning one or more gates point by point, recording the signal
$U_\text{fdbk}$ (or $U_\text{out}$) for typically 1~s and performing statistics
and Fourier transform. The gate voltage step size could be set arbitrarily
small, while the resolution was limited by thermal or vibrational noise of
the signal lines ($\sim 10^{-8}$~V rms).

The readout electronics for this experiment was optimized for stability
rather than sensitivity or band width in order to minimize external disturbances
on the SET and its environment.

\subsection{\label{sec:results}Results}
SET devices of type A and B (see Fig.~\ref{layout}) were systematically
investigated by taking $IV$ characteristics and noise figures as a function of
the four gates. The individual gate capacitances, which are listed in
Table~\ref{tab:devices}, were consistently determined by measurement of Coulomb
blockade peak spacings $\Delta U_\text{G} = e / C_\text{G}$ as well as from
the feedback method using Eq.~(\ref{eq:fdbk}).

\begin{table}[h]
\caption{\label{tab:devices}Gate capacitances in units of (aF) for two
multigate SETs (see Fig.\ref{layout}) as determined from Coulomb blockade peak
spacing measurements.}
\begin{ruledtabular}
\begin{tabular}{crrrr}
device type & G1 & G2 & G3 & G4 \\
A & 25.4 & 9.89 & 3.30 & 9.63 \\
B & 28.3 & 5.40 & 5.54 & 3.76 \\
\end{tabular}
\end{ruledtabular}
\end{table}

By means of the feedback circuitry the signal $U_\text{fdbk}$ was measured as a
function of the voltages varied at one or more gates. We define the
``background charge offset signal''
\begin{equation}
\Delta Q_\text{offset} = \left(\Delta U_\text{fdbk} +
\Delta U_{i} \frac{C_{i}}{C_\text{fdbk}}\right) C_\text{fdbk},
\label{eq:qoffset}
\end{equation}
where $\Delta U_{i}$ is the step size of the gate sweep.  $Q_\text{rms}$ will
then be denoting the rms noise on the $\Delta Q_\text{offset}$ signal.
The value of $\Delta Q_\text{offset}$ is zero when the SET electrometer
is linearly modified by the
gate potentials only, and shows peaks for discontinuous variations of background
charges.  By scanning one single gate up and down, we found non-zero $\Delta
Q_\text{offset}$ events which reproducibly appear at the same gate voltage
$U_{i}$ while showing hysteresis of different magnitude. The observations are
consistent with former measurements on background charge
fluctuations.~\cite{cbosc.r} However, in Ref.~\onlinecite{cbosc.r} only a
discrete resolution of the gate dependence was possible via analysis of the
Coulomb blockade peak positions, whereas here we are able to scan an arbitrary
and continuous gate range.

\subsubsection{Sweeping of two gates}\label{subsubsec:2gates}
\begin{figure}[h]
\includegraphics[width=8.6cm]{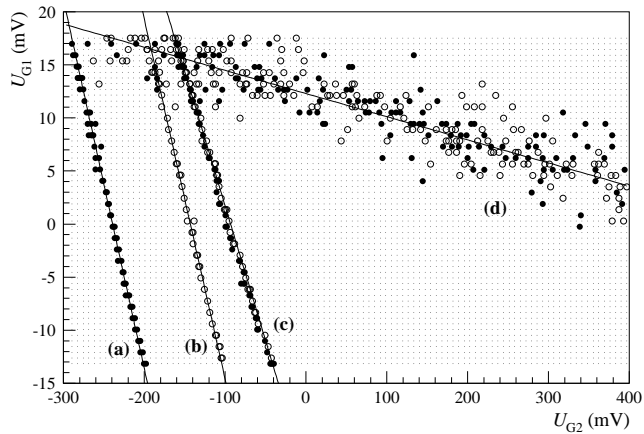}
\caption{\label{fig:scan2d} Example of background charge fluctuation events with
$(Q_\text{rms})^2 + (\Delta Q_\text{offset})^2 > (0.03 e)^2$, due to
variation of gates G2 and G1, as measured with a type A device. The straight
lines are numerically fitted to the data. Circles and dots are drawn for up and
down sweeps of $U_\text{G2}$, respectively. Only every fifth measurement point
(small dots) in $U_\text{G2}$ is shown for clarity.}
\end{figure}

If two gates are swept simultaneously (plus a third gate always acting as
feedback), the local potential around the multigate SET is differently
modified due to the device geometry. The second gate shifts the potential
threshold of a charge trap depending on its position. A measurement with two
simultaneous gates swept is shown in Fig.~\ref{fig:scan2d}. The circles and dots
correspond to events from up and down sweeps of gate G$_2$, respectively, where
$|\Delta Q_\text{offset}|$ is above the instrumental and background noise. After
each G$_2$ sweep, $U_\text{G1}$ is incremented. The events fall on different
straight lines with different slopes. Depending on the linear combination of the
gate voltages involved, patterns different from Fig.~\ref{fig:scan2d} were
obtained, i.e.\ including positive slopes, displaying more or less scatter,
showing merging event lines or sudden breaks. Measurements with all possible
gate permutations on the same device were performed.

Under constant cryogenic conditions, the experimental results were highly
reproducible, even after weeks. Although the noise spectrum level of our devices
($10^{-4} \dots 10^{-3} e/\sqrt{\text{Hz}}$ at 10~Hz) is
comparable to literature values for standard SETs,~\cite{setnoise.r}
we mostly observe a remarkably high long-term stability of our samples at stable
conditions, i.e.\ no drift and hardly any spontaneous jumps in $IV$
characteristics. We attribute this high stability in first line to the very slow
cooling of the devices (about one day with very little exchange gas), which
allows the traps to equilibrate in their lowest states. However, it is
empirically established that the characteristics of background fluctuations
change drastically upon thermal cycling.

\begin{figure}[h]
\includegraphics[width=8.7cm]{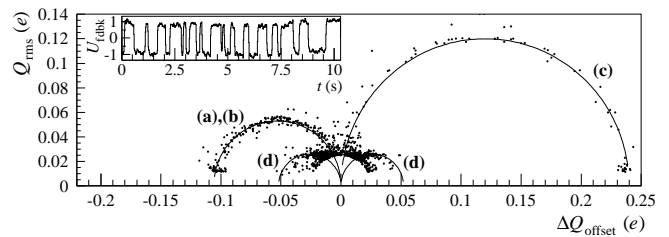}
\caption{\label{fig:s2n} Noise vs. signal in units of the elementary
charge $e$ plotted from data of the same measurement as in
Fig.~\ref{fig:scan2d}. The semi-circles are fitted to the data points as
described in the text. The intrinsic noise with $Q_\text{rms}^2 + \Delta
Q_\text{offset}^2  < (0.025 e)^2 $ is cut off for clarity.
The inset shows a $U_\text{fdbk}$ signal (in arbitrary units) as a function of
time at a transition point where increased two-level fluctuations occur.}
\end{figure}

\subsubsection{Signal height}\label{subsubsec:sigheight}
Signal height $\Delta Q_\text{offset}$ and rms noise $Q_\text{rms}$ from one
measurement are shown in Fig.~\ref{fig:s2n}, taking the same data as in
Fig.~\ref{fig:scan2d}. A clear semi-circle correlation between noise and signal
is observed. The lines are fits of
\begin{equation}
(|\Delta Q_\text{offset}| - \frac{|\Delta Q_0|}{2})^2 + {Q_\text{rms}}^2 =
(\frac{\Delta Q_0}{2})^2
\label{circle1.eq}
\end{equation}
to the data points, where $\Delta Q_0$ is the full signal height due to one
individual trap transition. This correlation is a consequence of the
measurement system: $Q_\text{rms}$, which is the rms value of the signal taken
typically for 1~s, averages quadratically over possible two-level fluctuations.
Assuming that the signal fluctuates during the measurement between zero and
$\Delta Q_0$ where it spends the total times $t_1$ and $t_2$, respectively, and
the transitions are negligibly fast, the average and the rms values of
this signal would be
\begin{equation*} \Delta Q_\text{offset} = \frac{t_2 \Delta Q_0}{t_1 + t_2} ,
\end{equation*}
\begin{equation}
Q_\text{rms} = \sqrt{ \frac{ t_1 \Delta Q_\text{offset}^2 +
t_2 (\Delta Q_0 - \Delta Q_\text{offset})^2 } {t_1 + t_2}}.
\label{circle2.eq}
\end{equation}
It is easily shown that the relations~(\ref{circle2.eq})  correspond to
Eq.~(\ref{circle1.eq}). One interesting information from Fig.~\ref{fig:s2n} is
the correspondence between the events falling on a semi-circle and those
building a straight line in Fig.~\ref{fig:scan2d}, which is represented by the
labels (a)-(d). This allows to identify and correlate switching events from
individual traps by both the gate dependence and the signal height.

One might expect that the data in Fig.~\ref{fig:s2n} yield detailed information
on capture and escape rates of the traps involved. However, we have found no
correlation between $Q_\text{rms}$ (or $\Delta Q_\text{offset}$) and the
position relative to the events on the lines in Fig.~\ref{fig:scan2d}. This is
due to metastability of the states where time constants are exponentially
large, as discussed in section~\ref{sec:discuss}.

\subsubsection{Gate polarization and noise}\label{subsubsec:noise}
\begin{figure}[h]
\includegraphics[width=8.6cm]{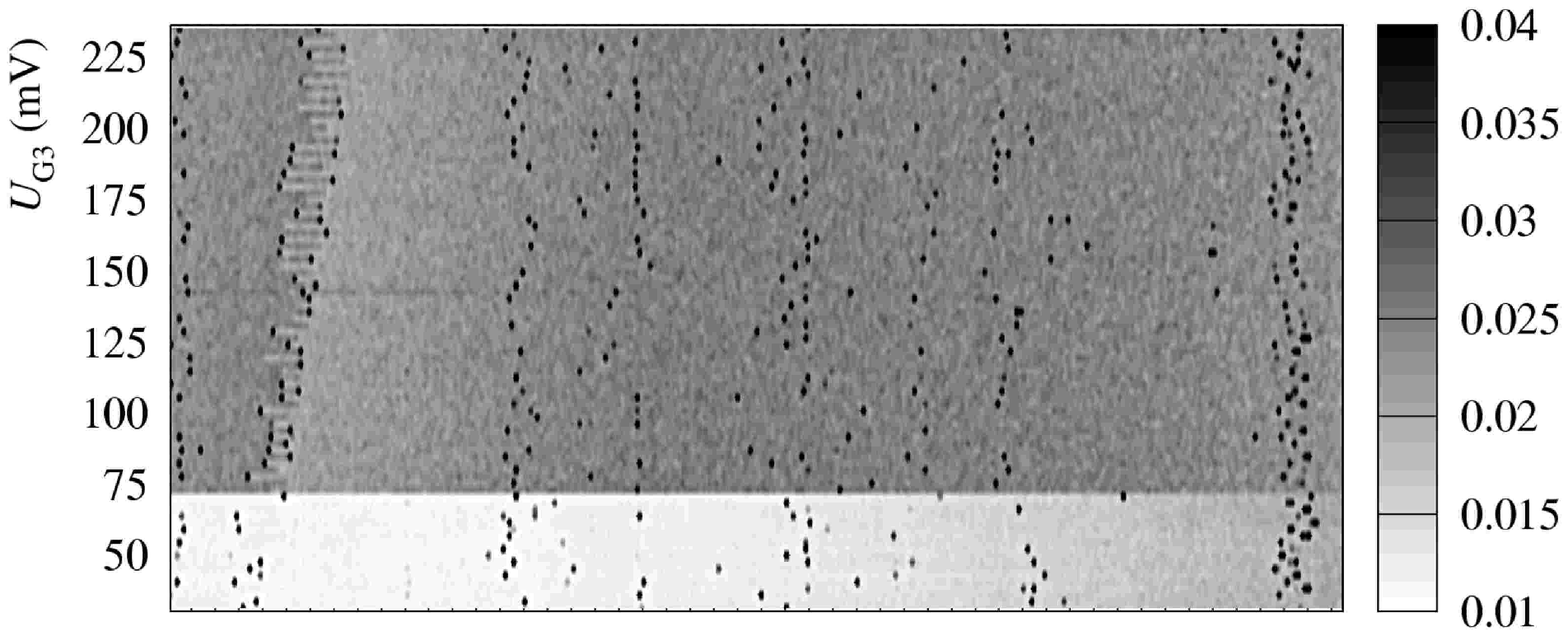}
\includegraphics[width=8.6cm]{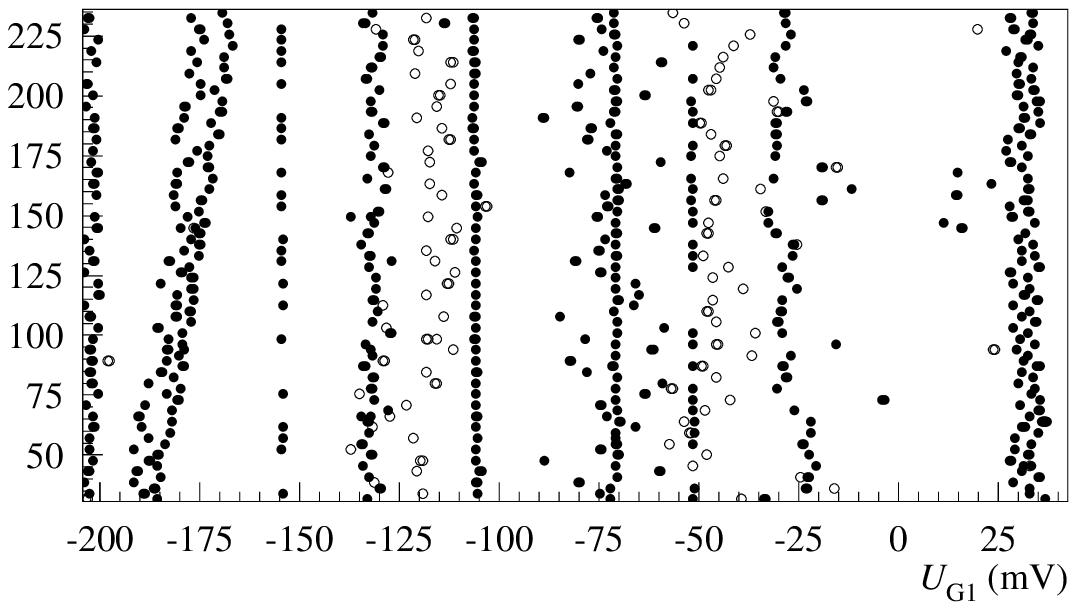}
\caption{\label{fig:noiselev} Top: Gray scale plot of noise level $Q_\text{rms}$
of the feedback signal (in units of $e$, see scale on the right) as a function
of the gate voltages $U_\text{G1} , U_\text{G3}$ measured on a type B device
(experimentally, $U_\text{G3}$ is incremented after each consecutive
$U_\text{G1}$ scan).
The data are compressed at $1/3$ of the full
noise spectrum to enhance visibility.
Bottom: Background charge switching events for $\Delta Q_\text{offset} > 0.2
e$ (same data as above). Circles and dots are drawn for up and down gate sweeps,
respectively. Signals with negative polarity are omitted for clarity.}
\end{figure}

The noise figures of SETs are in general determined by input charge noise and
the level is proportional to the SET gain.~\cite{starmark.r} However, in our
case where the operation point (i.e. gain, island charge) is kept constant,
eliminating the dominant noise dependence, we observe the noise level from
background charge fluctuations to clearly depend on the polarization of the
surrounding by variation of the gate potentials, as shown in
Fig.~\ref{fig:noiselev}.

The large fluctuations (black dots) in the upper part of Fig.~\ref{fig:noiselev}
naturally correspond to the events drawn in the lower part where
$\int{\Delta Q_\text{offset} \, \text{d}U_\text{G1}}$ experiences a step
response at the transition point and $Q_\text{rms}$ averages over the two level
states. In addition, a significant abrupt change in noise is observed at
$U_\text{G3} = 73$~mV and in the range $-190\text{~mV} < U_\text{G1} <
-170\text{~mV}$. Also gradual variations of the noise level as a function of gate
potential are measurable. While the onset of higher noise at $U_\text{G3} =
73$~mV is observed as a unique event with increasing $U_\text{G3}$, the features
around $U_\text{G1} \approx -180$~mV show an interesting combination of noise
figure and trap metastability: with increasing $U_\text{G1}$ the trap alters its
charge configuration such that the charge noise seen by the SET is reduced. By
reversing the G1 sweep direction, the noise level is again minimal right before
the trap switches back, whereas the gate voltage threshold of that transition is
lower.

The transition at $U_\text{G3} = 73$~mV in the upper plot of Fig.~\ref{fig:noiselev}
does not show  switching events in the lower plot. This is, besides an apparent
insensitivity of that trap to $U_\text{G1}$, an artefact of the
measurement procedure: $U_\text{G3}$ is incremented at the end of an $U_\text{G1}$
sweep, therefore transitions dominated by $U_\text{G3}$ occur only at
the turning points, i.e.\ the $U_\text{G1}$ extrema, and are not drawn in the plots.

\subsubsection{Superconductivity and magnetic field}\label{subsubsec:supercond}
While most of the measurements have been performed with the SET in the normal
state, we have also carried out test experiments in the superconducting
state to investigate reproducibility. In none of the cases have we found any
difference in feedback response between normal and superconducting devices. Nor
did we find any indication of magnetic field dependence (for normal state SET
with $B > 0.5$~T).

\subsubsection{No bias voltage dependence}\label{subsubsec:vbias}
A further important observation is the independence of the experimental results
from the biasing conditions, i.e.\ the source-drain voltage, neither on
magnitude nor on polarity. This has been reported already in
Ref.~\onlinecite{cbosc.r}. The electric field within a thin tunnel
junction changes drastically upon small variation of the source-drain voltage,
which should yield a clearly observable effect in case of trap
polarization inside the junction. In contrast, the surrounding is hardly
influenced for comparable voltages ranging within the Coulomb gap
(on the order of 1~mV).

\subsubsection{Spread of Coulomb blockade peak spacings}
\label{subsubsec:cbspace}
Finally, it's worth noticing an interesting experimental fact we
additionally discovered during measurements: when accurately
determining the Coulomb blockade peak positions and calculating the
nearest neighbor peak spacings $\Delta U_{G}$ (which are
theoretically expected to be exactly $e/C_{G}$), we observed an
increased statistical variation of $\Delta U_{G}$ for measurements
with gates at larger distance, i.e\ lower $C_{G}$. This is at first
sight counter intuitive, because  voltage noise $U_{n}$ (from readout
electronics or noise on the cables) at the gates with lower
$C_{G}$ should have much less effect on the SET island than nearby
gates, since the image charge influenced on the island is
proportional to $U_{n}C_{G}$. A possible interpretation of the results is
related to the trap density of states between island and gate. The
larger the number of charge fluctuators polarizable by the gate
variations, the larger the statistical fluctuations sensed by the SET
island.
The measurements can be considered as a statistical probe of the charged
environment in contrast to the experiments with discrete charge switching
events. And because a larger voltage variation at more distant gates is required
to modulate the SET island charge by one electron, more traps in those gate regions
experience a transition, which is reflected by increased statistical (not amplitude)
fluctuations.

Although the phenomenon is clearly, consistently and reproducibly
observed, it is unfortunately a very small effect (a few percent)
which doesn't permit quantitative analysis yet. More experiments with
specifically designed structures may improve resolution and allow
e.g.\ determination of a trap density of states.

\section{\label{sec:discuss}Discussion}
The invariability of our results upon different biasing conditions (source-drain
voltage, superconducting device) mentioned right above strongly suggests that
the switching processes observed can not reside within the tunnel junctions
themselves. Apart from a few exceptions in literature~\cite{tunneltrap.r} where
the results were interpreted in terms of trap fluctuations within the tunnel
junction, it is generally believed that the dominant low frequency noise source
is due to fluctuators located within the substrate or the oxide surrounding the
SET device.~\cite{setnoise.r}

Taking this into account, the simplest model to describe the process responsible
for the observed charge switching events is a bistable
trap,~\cite{tlfrev.r} located somewhere
close to the SET island, where a charge can hop (or rather tunnel) between two
adjacent sites back and forth.~\cite{olaf.r} This picture is equivalent to a
reversible dipole. Phenomenologically the polarity of the charge is in our case
of no importance and isn't distinguishable either.
Therefore the charge carrier involved is always taken
to be an electron for simplicity. The trap configuration can change by
variation of potentials (i.e.\ gate voltages), by thermal activation or due to
PAT. The results from our experiments are clearly
dominated by the potential landscape intentionally modified by the four gates,
although other effects may play a role at much smaller energy scales.

For illustration imagine an electron trapped in a local impurity state.
An electric field, which is in our case determined by linear
superposition of the potentials of the four gates and the
source-island-drain structure, can force that electron to switch to a
nearby metastable trap site. If the electron e.g.\ shifts closer to the SET
island, the latter experiences a charge reduction, which is
compensated by a signal from the feedback circuit according to
Eq.~(\ref{eq:qoffset}). This explains geometrically the signal height
and polarity of individual fluctuators as presented in
section~\ref{subsubsec:sigheight}.

The striking reproducibility of switching events as a function of the gate
voltages allows detailed analysis of the traps involved. The data shown in
Figs.~\ref{fig:scan2d} and \ref{fig:s2n} nicely illustrate the
distinct nature of individual traps. From the lines fitted to the
data in Fig.~\ref{fig:scan2d} it was derived that the events of (a) and (b) have
exactly the same slopes $\text{d}U_\text{G1} / \text{d}U_\text{G2}$, and they
also show the same signal height spectrum in Fig.~\ref{fig:s2n}. They are only
distinguished by their transition voltage threshold, which depends on gate sweep
direction with a specific hysteresis of $\Delta U_\text{G2} =
93.2$~mV in this case. According to
the trapping process (charging--discharging) we like to refer to the events
represented by circles and dots as flips and flops, respectively. In contrast to
the events (a) and (b), the transition thresholds of flips and flops on
line (c) are
indistinguishable, i.e.\ displaying a vanishing hysteresis. Yet another behavior
is observed for the events on (d) which scatter very strongly. They also show
both signal polarities (but of same magnitude), irrespective of gate sweep
direction (see Fig.~\ref{fig:s2n}). This specific trap is unstable within an
unusually broad gate voltage window. This  is probably caused by
interactions with other traps in close vicinity, rather than by pure thermal
activation or PAT which we can exclude due to the large
excitation energies required.

\begin{figure}[h]
\includegraphics[width=8.6cm]{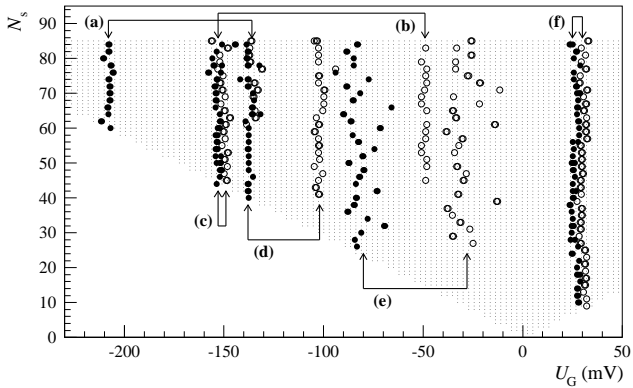}
\caption{\label{fig:triangle} Charge fluctuations with $\Delta Q_\text{offset}
\geq 0.02 e$ measured with a type B device by increasing the gate scan range
with increasing scan number $N_\text{s}$ (only a selected part of the entire
measurement is shown). Events with negative $\Delta Q_\text{offset}$ polarity
show similar correlations, but they are omitted for clarity. The flip and flop
thresholds according to transitions of one individual trap can be related by
their onset and are denoted by arrows. The small dots (grid) represent every fifth
measurement point in $U_\text{G}$.}
\end{figure}

In addition to the feasibility of assigning the experimental data as shown in
Figs.~\ref{fig:scan2d} and \ref{fig:s2n} to individual traps, we can also make
an identification by relating the flip and flop thresholds of the same trap from
an experiment as presented in Fig.~\ref{fig:triangle}.

The measurement in Fig.~\ref{fig:triangle}, where the gate voltage range is
increased with each consecutive sweep, clearly shows that a trap can only
switch off (flop) if it has also been switched on (flip) before, or vice versa.
This is consistent with and suggestive of the single charging nature of the
traps observed here.

Although some events in Fig.~\ref{fig:triangle} are detected at the same
threshold, i.e.\ the flops of (b) and (c) or the (a) flips and the (d) flops,
we can again easily identify them due to clear distinction in the noise vs.\
signal height diagram (like in Fig.~\ref{fig:s2n}, but not shown in this case).

The behavior of the traps in general seems to be only very weakly influenced by
other fluctuators. This is concluded from the straight lines found in
measurements like in Figs.~\ref{fig:scan2d} or \ref{fig:triangle}. The
transition threshold is usually not altered irrespective of the polarization
strength of the environment
(Fig.~\ref{fig:triangle}), i.e.\ there is no difference
whether a small or a large gate potential was applied,
the trap switching occurs at the same gate voltages.
A rarely clear exception is seen
in the events (f) in Fig.~\ref{fig:triangle}, which show slight shifts at the
onsets of traps (a) and (e), i.e.\ for $N_\text{s} = 60$
and 26, respectively.
This suggests a rather small distance of
these particular traps among each other
allowing electrostatic interaction. In general, however, according
to simple electrostatics,
charges (or traps) in very close vicinity to metallic electrodes are strongly
screened.~\cite{sitraps.r} The suppressed interaction leads to the
observation of individual unperturbed charge trapping events.

From the characteristics of individual traps, i.e.\ their transition thresholds
as a function of polarization of their environment, we can not only
distinguish but also localize  the fluctuators. Variation of
two gates electrostatically determines the equipotential line where
the trap switching occurs. The relation of the gate potentials is
extracted from the slope
$\text{d}U_{\text{G}i} / \text{d}U_{\text{G}j}$
of an individual trap as taken e.g.\ from Fig.~\ref{fig:scan2d}.
Addition of more gates (as are actually present in our experiments with
the feedback gate and such with static potential) simply modifies the
linear combination of electric potentials involved, allowing to
verify the results from experiment and calculations. The exact
position (and size) of the trap relative to the SET island is then determined
by the signal height and polarity. An estimate for charge variations
on the island on the order of $0.1 e$ shows that the observable traps
(with a typical extension of $\approx 2$~nm) must
reside in the very close vicinity of the island.~\cite{cbosc.r}

\begin{figure}[h]
\includegraphics[width=8.6cm]{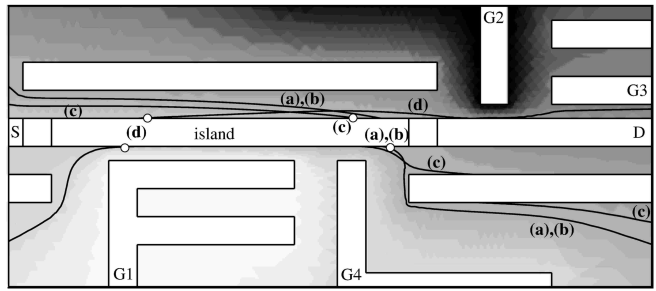}
\caption{\label{fig:efield}  The potential landscape (gray scale) is shown for
the slopes (a,b) from Fig.~\ref{fig:scan2d}. Calculated equipotential lines for
zero variation are drawn for all slopes from that experiment. The white dots at
the edge of the island mark the positions of the fluctuators. Accuracy
of the calculated position tangential to the island
is numerically estimated to be about 10~nm. The
perpendicular position is given by the signal height and limited to a
few nanometers. }
\end{figure}

We have performed numerical electrostatics calculation (equipotential
lines) on the device, based on a quasi-threedimensional finite
element method (real twodimensional finite element calculations in
the substrate plane with iterative corrections
in the third dimension). An example of a potential calculation
for the trap events (a) and (b) of  Fig.~\ref{fig:scan2d} is shown in
Fig.~\ref{fig:efield}. Gates G3, G4 and source-island-drain are at
zero potential while gates G1 and G2 are at relative potentials
defined by the slope $\text{d}U_\text{G1} / \text{d}U_\text{G2}$ of
(a,b). This yields an equipotential line for zero variation labelled
(a),(b). The white dot
(on the lower right edge of the island, also labelled (a),(b))
finally represents the location of that particular trap.
Calculated equipotential lines are also drawn for trap events (c) and
(d), respectively. The trap location (d) is ambiguous and drawn on
both island sides due to not well defined signal polarity (see
Fig.~\ref{fig:s2n}).
\footnote{Note that the equipotential lines seem to cross the island, which of
course is not possible with a metallic island in two dimensions. However, keep
in mind that Fig.~\ref{fig:efield} shows a three dimensional projection on the
2D plane (or a cut of equipotential surfaces).
The lines are effectively neither touching nor crossing the metal.
The trap positions are determined where the lines approach the island at a
distance of $\sim 2$~nm.}

This intriguing feature to determine the position of individual traps
while studying their specific behavior
opens up a new method to study the substrate,
surface, interface or oxide quality which are of increasing
importance in nanostructure devices.
Of course, the method is limited by the ability to observe distinct
finite slopes in $\text{d}U_{\text{G}i} / \text{d}U_{\text{G}j}$. The
majority of the data in Fig.~\ref{fig:noiselev} is therefore not
useful for trap position determination. The data simply reflects a
negligible influence of gate G3 on the measured traps due to screening
by other gates in our structure not specifically intended for such
experiments. The problem can easily
be resolved by optimized and more sophisticated lithographic design
(e.g.\ star-like arrangement of gates to minimize electrostatic screening
between traps and SET  island).

On the other hand we wish to emphasize the possibility to optimize the
operation point of a real device like an electron pump or a turnstile
by noise analysis as shown in section~\ref{subsubsec:noise}. Stable
operation of SET devices with respect to background charge fluctuations
is one of the biggest issues in their implementation into useful
circuits. With an appropriate device structure and the method
described here we can find the working point of optimum signal to
noise ratio or maximize device stability, as required for low noise
applications.

The task of hysteresis as observed in our experiments is rather
delicate and not fully understood. In general, hysteresis or
metastability may be due to asymmetric tunnel barriers, polarization
of the proximal environment (including other traps) or multiple
tunneling processes.~\cite{multijunction.r}
We are not able to decide on the dominant
microscopic origin of the hysteresis measured in our case. However,
our observation of increased fluctuations close to a transition point
indicates that there are different energy scales and more than one
single mechanism involved.
The time constants of the generation-recombination
processes~\cite{srh.r} are unfortunately, due to
hysteresis, either astronomically large, or at scales which are not
accessible with our experimental setup
(readout electronics bandwidth $<10$~kHz).
The spectrum of hysteresis values (energy
difference between flip and flop) was found to be flat up to several
100~mV (i.e.\ covering the entire gate sweep range). This prevents us
to extract quantitative data on the metastability of the traps which
apparently have an average energy spread much larger than is
accessibly with our method. This is also related to the fact that we do
not observe periodic multiple charging of the same trap. Probably the
trap size and the equivalent capacity are so small that the charging
energy can be far above 1~eV.

Nevertheless, we can make a very rough estimate of the trap density
assuming that all the observed fluctuators reside on the island
surface (i.e.\ traps in the oxide covering the island). The island
surface facing the gate (i.e.\ perpendicular to the electric field)
has an area of $60 \times 1000$~nm. We measure an extrapolated
average of 20 traps per volt of gate variation. That yields an
interface (or surface) trap density of 330~eV$^{-1} \mu$m$^{-2}$ or
roughly $3 \times 10^{10} {\text{eV}}^{-1} {\text{cm}}^{-2}$.
We believe this is  quite a reasonable number and
it is  well consistent with literature values~\cite{flicker.r}
(although they are given for Si, the order of magnitude should be
applicable).

As a final remark, we wish to comment on one possible origin of the
charge traps responsible for the switching events. Empirically we have
observed significantly less fluctuators in a `fresh' device at the
very first cool-down compared to measurements after thermal cycling.
This may suggest that the imperfections are (also) a consequence of
mechanical stress. The effect of electrical stress on noise
characteristics has been reported in Ref.~\onlinecite{cobdenstress.r}.

\section{Conclusion}
We have presented a new experimental method and measurement results
on electrometric performance of a modified single--electron device.
The technique allowed to study charge traps individually, exploring the
very fluctuators which are responsible for noise and instabilities,
severely degrading the operation of mesoscopic electronic devices.
In particular, the charge background was investigated by scanning
different gates simultaneously and recording abrupt changes of the
SET island polarization. The experimental results were highly
reproducible and therefore allowed extensive systematic studies.
We were able to distinguish switching
processes due to individual charge traps. The knowledge of trap
transition threshold and signal height allowed to geometrically
determine the trap position, which was typically not more than a few
nanometers away from the island surface. The tangential accuracy of the
analysis was estimated to be about 10~nm. This provides an alternative
method to e.g.\ scanning imaging experiments~\cite{trapimaging.r} for local
investigation of trapping processes.

The noise level of the SET device was found to depend on the 
environment polarization due to gate potential variations. The 
reproducible experiments offer a method to determine the noise 
characteristics of the SET  and to adjust the operation point 
for maximum signal to noise or best device stability. It is 
particularly advantageous that the testing of the charge background 
is performed with the same working device, i.e.\ optimization detector 
and subsequent application device are the same.

The fluctuators observed in the experiments are exclusively located 
in the close vicinity of the SET island, i.e.\ in the oxide covering
the island or in the substrate. In the former case we have roughly estimated a 
trap density of  $3 \times 10^{10} {\text{eV}}^{-1} {\text{cm}}^{-2}$, 
which compares very well with literature values.

\begin{acknowledgments}
We are grateful to Klaus Ensslin, Alexander B.~Zorin, Blaise Jeanneret,
Beat Jeckelmann and Felix Meli for valuable discussions and support.
\end{acknowledgments}

\end{document}